# The Possibilist Transactional Interpretation and Relativity

http://www.springerlink.com/content/6t61058t2m268855/?MUD=MP

Ruth E. Kastner

Foundations of Physics Group

UMCP

22 June 2012

ABSTRACT. A recent ontological variant of Cramer's Transactional Interpretation, called "Possibilist Transactional Interpretation" or PTI, is extended to the relativistic domain. The present interpretation clarifies the concept of 'absorption,' which plays a crucial role in TI (and in PTI). In particular, in the relativistic domain, coupling amplitudes between fields are interpreted as amplitudes for the generation of confirmation waves (CW) by a potential absorber in response to offer waves (OW), whereas in the nonrelativistic context CW are taken as generated with certainty. It is pointed out that solving the measurement problem requires venturing into the relativistic domain in which emissions and absorptions take place; nonrelativistic quantum mechanics only applies to quanta considered as 'already in existence' (i.e., 'free quanta'), and therefore cannot fully account for the phenomenon of measurement, in which quanta are tied to sources and sinks.

1. Introduction and Background

The transactional interpretation of quantum mechanics (TI) was first proposed by John G. Cramer in a series of papers in the 1980s (Cramer 1980, 1983, 1986). The 1986 paper presented the key ideas and showed how the interpretation gives rise to a physical basis for the Born Rule, which prescribes that the probability of an event is given by the square of the wave function corresponding to that event. TI was originally inspired by the

Wheeler-Feynman (WF) time-symmetric theory of classical electrodynamics (Wheeler and Feynman 1945, 1949). The WF theory proposed that radiation is a time-symmetric process, in which a charge emits a field in the form of half-retarded, half-advanced solutions to the wave equation, and the response of absorbers combines with that primary field to create a radiative process that transfers energy from an emitter to an absorber.

As noted in Cramer (1986), the original version of the Transactional Interpretation (TI) already has basic compatibility with relativity in virtue of the fact that the realization of a transaction occurs with respect to the endpoints of a space-time interval or intervals, rather than at a particular instant of time, the latter being a non-covariant notion. Its compatibility with relativity is also evident in that it makes use of both the positive and negative energy solutions obtained from the Schrödinger equation and the complex conjugate Schrödinger equation respectively, both of which are obtained from the relativistic Klein-Gordon equation by alternative limiting procedures. Cramer has noted in (1980, 1986) that in addition to Wheeler and Feynman, several authors (including Dirac) have laid groundwork for and/or explored explicitly time-symmetric formulations of relativistic quantum theory with far more success than has generally been appreciated.[1]

A modified version of TI, 'possibilist TI' or PTI, was proposed in Kastner (2010) and elaborated in Kastner (2011b), wherein it was shown that certain challenges mounted against TI can be satisfactorily addressed and resolved. This modified version proposes that offer and confirmation waves (OW and CW) exist in a sub-empirical, pre-spacetime realm (PST) of possibilities, and that it is actualized transactions which establish empirical spatiotemporal events. PST is considered to be the physical, if unobservable, referent for Hilbert Space (and, at the relativistic level, Fock Space). This paper is devoted to developing PTI in terms of a quantum relativistic extension of the Wheeler-Feynman theory by Davies (1970,71,72).

1.1 Emission and absorption are fundamentally relativistic processes.

---

[1] E.g., Dirac (1938), Hoyle and Narlikar (1969), Konopinski (1980), Pegg (1975), Bennett (1987).

It should first be noted that the concept of coupling is important for understanding the process of absorption in TI, which is often misunderstood. Under TI, an 'absorber' is an entity that generates confirmation waves (CW) in response to an emitted offer wave (OW). The generation of a CW needs to be carefully distinguished from 'absorption' meaning the absorption of energy, since not all quantum absorbers will in fact receive the energy from a given emitter. In general, there will be several or many absorbers sending CW back to an emitter, but only one of them can receive the emitted energy. This is purely a quantum effect, since the original classical Wheeler-Feynman absorber theory treats energy as a continuous quantity that is distributed to all responding absorbers. It is at the quantum level that a semantic difficulty arises, in that there are entities (absorbers) that *participate* in the absorption process by generating CW, but do not necessarily end up receiving energy. In everyday terms, these are like sweepstakes entrants that are necessary for the game to be played, but who do not win it.

A longstanding concern about the basic TI picture has been that the circumstances surrounding absorption are not well-defined, and that 'absorber' could therefore be seen as a primitive term. This concern is squarely addressed and resolved in the current approach as follows. PTI can indeed provide a non-arbitrary (though not deterministic) account for the circumstances surrounding absorption in terms of coupling between fields. Specifically, I propose that 'absorption' simply means annihilation of a quantum state, which is a perfectly well-defined physical process in the relativistic domain. Annihilation is defined by the action of an annihilation operator on an existing quantum state; e.g., $a_p |p> = |0>$. Meanwhile, the bra $<p|$, representing a CW, is created by the action of an annihilation operator on the vacuum state bra $<0|$.[2] Any measurement depends crucially on these processes, which are not representable in nonrelativistic quantum theory. The fact that objections to TI can be resolved at the relativistic level underscores both (1) the ability of the basic transactional picture to accommodate relativity and (2) the necessity to include the relativistic domain to resolve the measurement problem.

---

[2] Emission is of course defined by the action of the corresponding creation operator on the vacuum state: $a^{\dagger}_p |0> = |p>$.

Thus, the crucial feature of TI/PTI that allows it to "cut the Gordian knot" of the measurement problem is that it interprets absorption as a real physical process that must be included in the theoretical formalism in order to account for any measurement result (or more generally, any determinate outcome associated with a physical system or systems). The preceding is a specifically relativistic aspect of quantum theory, since nonrelativistic quantum mechanics ignores absorption: it addresses only persistent particles. Strictly speaking, it ignores emission as well; there is no formal component of the nonrelativistic theory corresponding to an emission process. The nonrelativistic theory is applied only to an entity or entities assumed to be already in existence. In contrast, relativistic quantum field theory explicitly includes emission and absorption through the field creation and annihilation operators respectively; there are no such operators in nonrelativistic quantum mechanics.[3] Because the latter treats only pre-existing particles, the actual emission event is not included in the theory, which simply applies the ket $|\Psi\rangle$ to the pre-existing system under consideration. Under these restricted circumstances, it is hard to see a physical referent for the bra $\langle\Psi|$ from within the theory, even though it enters computations needed to establish empirical correspondence (such as amplitudes and probabilities). What PTI does is to 'widen the scope' of nonrelativistic quantum theory to take into account both emission and absorption events, the latter giving rise to the advanced state or bra $\langle\Psi|$. In this respect, again, it is harmonious with relativistic quantum theory.

---

[3] Technically, the Davies theory, which is probably the best currently articulated theoretical model for TI and which is discussed below, is a direct action (DA) theory in which field creation and destruction operators for photons are superfluous; the electromagnetic field is not really an independent entity. Creation and annihilation of photons is then physically equivalent to couplings between the interacting charged currents themselves, and it is the coupling amplitudes that physically govern the generation of offers and confirmations. The important point is that couplings between fields are inherently stochastic and so are the generations of OW and CW.

1.2 TI/PTI retains isotropy of emission (and absorption).

It should also be noted that the standard notion of emission as being isotropic with respect to space (i.e., a spherical wave front) but *not* isotropic with respect to time (i.e., that emission is only in the *forward* light cone) seems inconsistent, and intrinsically ill-suited to a relativistic picture, in which space and time enter on an equal footing (except, of course, for the metrical sign difference). The prescription of the time-symmetric theory for half the emission in the $+t$ direction and half in the $-t$ direction is consistent with the known fact that emission does not favor one spatial direction over another, and harmonious with the relativistic principle that a spacetime point is a unified concept represented by the four-vector $x^\mu = \{x^0, x^1, x^2, x^3\}$. This symmetry principle, and the consistency concern related to it, rather than a desire to eliminate the field itself (historically the motivation for absorber-based electrodynamics, see below), is the primary motivation for PTI in its relativistic application.

2. The Davies Theory

I turn now to the theory of Davies, which provides a natural framework for PTI in the relativistic domain.

2.1 Preliminary remarks

The Davies theory has been termed an 'action at a distance' theory because it expresses interactions not in terms of a mediating field with independent degrees of freedom, but rather in terms of direct interactions between currents.[4] As Cramer (1986) notes, one of the original motivations for such an 'action at a distance' theory was to eliminate troubling divergences, stemming from self-action of the field, from the standard theory; thus it was thought desirable to eliminate the field as an independent concept.

---

[4] The term 'current' in this context denotes the generalization of a probability distribution for a particle associated, in the relativistic domain, with a quantum field.

However, it was later realized that some form of self-action was needed in order to account for such phenomena as the Lamb shift (although the Davies theory does allow for self-action in that a current can be regarded as acting on itself in the case of indistinguishable currents (see, e.g., Davies (1971), 841, Figure 2)).

Nevertheless, despite its natural affinity for a time-symmetric model of the field, it must be emphasized that PTI does *not* involve an ontological elimination of the field. On the contrary, the field remains at the 'offer wave' level. This is the same picture in which the classical Wheeler-Feynman electromagnetic retarded field component acts as a 'probe field' that interacts with the absorber and prompts the confirming advanced wave, which acts to build up the emitter's retarded field to full strength and thus enable the exchange of energy between the emitter and the absorber.

Thus PTI is based, not on elimination of quantum fields, but rather on the time-symmetric, transactional character of energy propagation by way of those fields, and the assumption that offer and confirmation waves capable of resulting in empirically detectable transfers of physical quantities only occur in couplings between field currents. However, in keeping with this possibilist reinterpretation, the field operators and fields states themselves are considered as pre-spacetime objects. That is, they exist; but not in spacetime. What exist in spacetime are actualized, measurable phenomena such as energy transfers. Such phenomena are always represented by real, rather than complex or imaginary, mathematical objects. At first glance, this possibilist ontology may seem strangely 'ephemeral'. However, when one recalls that a standard expression of quantum field theory such as the vacuum state $|0>$ has no spacetime arguments and is maximally nonlocal,[5] it seems reasonable to suppose that the physical entity referred to by this quantity exists, but not in spacetime (in the sense that it cannot be associated with any well-defined region in spacetime).

---

[5] This is demonstrated by the Reeh-Schlieder Theorem; cf. Redhead (1995).

A further comment is in order regarding PTI's proposal that spacetime is emergent rather than fundamental. In the introductory chapter to their classic *Quantum Electrodynamics*, Beretstetskii, Lifschitz and Petaevskii make the following observation concerning QED interactions:

"For photons, the ultra-relativistic case always applies, and the expression *[Δq ~ ℏ/p ], where Δq* is the uncertainty in position] is therefore valid. This means that the coordinates of a photon are meaningful only in cases where the characteristic dimension of the problem are large in comparison with the wavelength. This is just the 'classical' limit, corresponding to geometric optics, in which the radiation can be said to be propagated along definite paths or rays. In the quantum case, however, where the wavelength cannot be regarded as small, the concept of coordinates of the photon has no meaning. ...

The foregoing discussion suggests that the theory will not consider the time dependence of particle interaction processes. It will show that in these processes there are no characteristics precisely definable (even within the usual limitations of quantum mechanics); *the description of such a process as occurring in the course of time is therefore just as unreal as the classical paths are in non-relativistic quantum mechanics*. The only observable quantities are the properties (momenta, polarization) of free particles: the initial particles which come into interaction, and the final properties which result from the process." [The authors then reference L. D. Landau and R. E. Peierls, 1930][6]. (Emphasis added.) " (Beretstetskii, Lifschitz and Petaevskii 1971, p. 3)

The italicized sentence asserts that the interactions described by QED (and, by extension, by other interacting field theories) cannot consistently be considered as taking place in spacetime. Yet they do take place *somewhere*; the computational procedures deal with entities implicitly taken as ontologically substantive. This 'somewhere' is just the pre-spatiotemporal, pre-empirical realm of possibilities proposed in PTI. The 'free particles' referred to in the last sentence of the excerpt exist within spacetime, whereas the virtual (unobserved) particles do not.

---

[6] The Landau and Peierls paper has been reprinted in Wheeler and Zurek (1983).

2.2 Specifics of the Davies Theory

The Davies theory (1970,71,72) is an extension of the Wheeler-Feynman time-symmetric theory of electromagnetism to the quantum domain by way of the S-matrix (scattering matrix). This theory provides a natural framework for PTI in the relativistic domain. The theory follows the basic Wheeler-Feynman method by showing that the field due to a particular emitting current $j_{(i)}^{\mu}(x)$ can be seen as composed of equal parts retarded radiation from the emitting current and advanced radiation from absorbers. Specifically, using an S-matrix formulation, Davies replaces the action operator of standard QED,

$$J = \sum_i \int dx\, j^{\mu}{}_{(i)}(x) A_{\mu}(x) \qquad (1)$$

(where $A_{\mu}$ is the standard quantized electromagnetic field), with an action derived from a direct current-to-current interaction,[7]

$$J = -\frac{1}{2} \sum_{i,j} \int dx\, dy\, j^{\mu}{}_{(i)}(x) D_F(x-y) j_{(j)\mu}(y), \qquad (2)$$

where $D_F(x-y)$ is the Feynman photon propagator. (This general expression includes both distinguishable and indistinguishable currents.)

---

[7] That these expressions are equivalent is proved in Davies (1971) and reviewed in (1972). The currents $j^{\mu}$ are fermionic currents.

While $D_F(x-y)$ implies a kind of asymmetry in that it only allows positive frequencies to propagate into the future, Davies shows that for a 'light-tight box' (i.e., no free fields), the Feynman propagator can be replaced by the time-symmetric propagator $\overline{D}(x) = \frac{1}{2}\left[D^{ret}(x) + D^{adv}(x)\right]$, where the terms in the sum are the retarded and advanced Green's functions (solutions to the inhomogeneous wave equation).

Specifically, Davies shows that if one excludes scattering matrix elements corresponding to transitions between an initial photon vacuum state and final states containing free photons, his time-symmetric theory, based on the time-symmetric action $J = -\frac{1}{2}\sum_{i,j}\int dx dy\, j^\mu{}_{(i)}(x)\overline{D}(x-y)j_{(j)\mu}(y)$, is identical to the standard theory. (See Davies 1972, eqs. (7-10) for a discussion of this point, including the argument that if one considers the entire system to be enclosed in a light-tight box, this condition holds.) The excluded matrix elements are of the form $\langle n|S|0\rangle$, where $n$ is different from zero. By symmetry, for emission and absorption processes involving (theoretically)[8] free photons in either an initial or final state, one must use $D_F$ instead of $\overline{D}$ to obtain equivalence with the standard theory.

To understand this issue, recall Feynman's remark that if you widen your area of study sufficiently, you can consider all photons 'virtual' in that they will always be emitted and absorbed somewhere.[9] He illustrated this by an example of a photon propagating from the Earth to the Moon:

---

[8] The caveat 'theoretically' is introduced because a genuinely free photon can never be observed: any detected photon has a finite lifetime (unless there are 'primal' photons which were never emitted) and is therefore not 'free' in a rigorous sense. This is elaborated below and in footnote 11.

[9] Feynman (1998). Sakurai ( 1973, 256) also makes this point: "As a matter of fact most real photons of physical interest are, strictly speaking, virtual in the sense that they are emitted at some place and absorbed at another place."

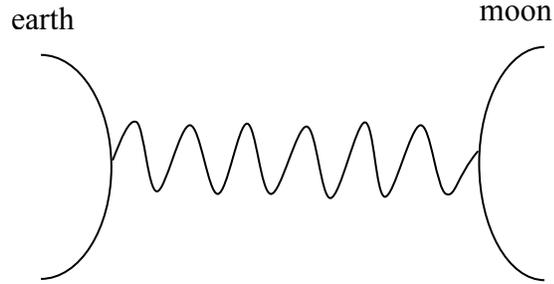

Figure 1. A "virtual" photon propagating from the earth to the Moon.

But, as Davies notes, this picture tacitly assumes that real (not virtual) photons are available to provide for unambiguous propagation of energy *from* the Earth *to* the moon. If such free photons are involved, then (at least at the level of the system in the drawing) we don't really have the light-tight box condition allowing for the use of $\overline{D}$ rather than $D_F$. (In any case, $\overline{D}$ alone would not provide for the propagation of energy in only one direction; time-symmetric energy propagation in a light-tight box in an equilibrium state would be fully reversible. Thus the observed time-asymmetry of radiation must always be explained by reference to boundary conditions, either natural or experimental). So one cannot assume that equivalence with the standard theory is achieved by the use of $\overline{D}$ for all photons represented by internal lines (i.e., for 'virtual' photons in the usual usage). One needs to take into account whether energy sources are assumed to be present on either end of the propagation. Thus, within the time-symmetric theory, the use of $D_F$ is really a *practical postulate,* applying to subsets of the universe and/or to postulated boundary conditions consistent with the empirical fact that we observe retarded radiation. It assumes, for example, that the energy source at the earth consists of 'free photons' rather than applying a direct-interaction picture in which the energy source photons arise from another current-current interaction and are therefore truly virtual.

The ambiguity surrounding this real vs. virtual distinction arises from the fact that a genuinely 'real' photon must have an infinite lifetime according to the uncertainty principle, since its energy is precisely determined at $k^2=0$.[10] But nobody will ever detect such a photon, since any photon's lifetime ends when it is detected, and the detected photon therefore has to be considered a 'virtual' photon in that sense. The only way it could truly be 'real' would be if it had existed since $t = -\infty$.[11] On the other hand, it is only detected photons that transfer energy; so, as Davies points out, photons that are technically 'virtual' can still have physical effects. It is for this reason that PTI eschews this rather misleading 'real' vs. 'virtual' terminology and speaks instead of offer waves, confirmation waves, and transactions—the latter corresponding to actualized (detected) photons. The latter, which by the 'real/virtual' terminology would technically have to called 'virtual' since they have finite lifetimes, nevertheless give rise to observable phenomena (e.g., energy transfer). They are contingent on the existence of the offer and confirmation waves that also must be taken into account to obtain accurate predictions (e.g., for scattering cross-sections, decay probabilities, etc.). So in the PTI picture, all these types of photons are real; some are *actualized* --a stronger concept than real--and some are just offers or confirmations. But since they all lead to physical consequences, they are all physically real, even if the offers and confirmations are sub-empirical (recall the discussion at the end of Section 1).

There is another distinct, but related, issue arising in the time-symmetric approach that should be mentioned. Recall (as noted in Cramer 1986) that a fully time-symmetric approach leads to two possible physical cases: (i) positive energy propagates forward in time/negative energy propagates backward in time or (ii) positive energy propagates backward in time/negative energy propagates forward in time. Thus the theory underdetermines specific physical reality.[12] We are presented with a kind of 'symmetry breaking': we have to choose which theoretical solution applies to our physical reality. In cases discussed above, in which fictitious 'free photons' are assumed for convenience, the use of $D_F$ rather than its inverse $D_F^*$ constitutes the choice (i). While this might be seen as grounds to claim that PTI is not 'really' time-symmetric, that judgment would not be valid, because it could be argued that what is considered 'positive' energy is merely conventional. Either choice would lead to the same empirical phenomena; we would merely have to change the theoretical sign of our energy units.

---

[10] "Off-shell" behavior applies in principal for any photon that lacks an infinite lifetime; this is expanded on in § 3.5.

[11] Of course, this is theoretically possible (even if not consistent with current 'Big Bang' cosmology), and could be regarded as the initial condition that provides the thermodynamic arrow, as well as an interesting agreement with the first chapter of Genesis. But the existence of such 'primal photons' would not rule out the direct emitter-absorber interaction model upon which TI is based. It would just provide an unambiguous direction for the propagation of positive energy.

3. PTI applied to QED calculations

3.1 Scattering: a standard example

In nonrelativistic quantum mechanics, one is dealing with a constant number of particles emitted at some locus and absorbed at another; there are no interactions in which particle type or number can change. However, in the relativistic case, with interactions among various coupling fields, the number and type of quanta are generally in great flux. A typical relativistic process is scattering, in which (in lowest order) two 'free' quanta interact through the exchange of another quantum, thereby undergoing changes in their respective energy-momenta $p$. A specific example is Bhabha (electron-positron) scattering, in which two basic lowest-order processes contribute as effective 'offer waves' in that they must be added to obtain a final amplitude for the overall process. The following two Feynman diagrams apply in this case:

---

[12] While this might seem as a drawback at first glance, the standard theory simply disregards the advanced solutions in an *ad hoc* manner (which, as noted previously, is inconsistent with the unification of space and time required by relativity). In the time-symmetric theory, the appearance of a fully retarded field can be explained by physical boundary conditions.

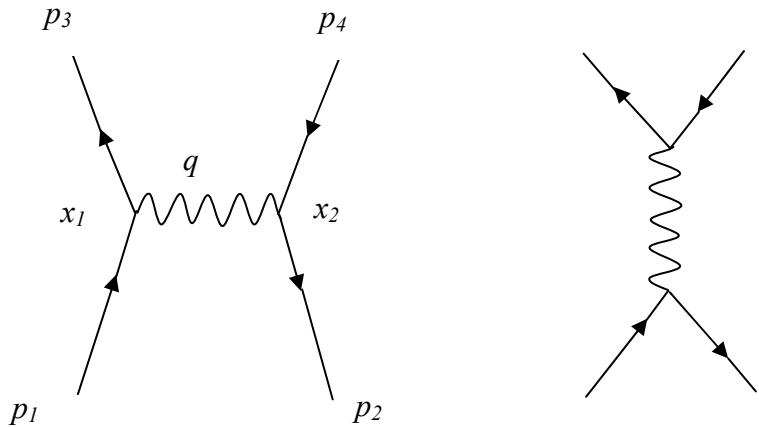

Figure 2. Bhabha scattering: the two lowest-order graphs.

For conceptual purposes I will discuss a simplified version of this process in which I ignore the spin of the fermions and treat the coupling strength (strength of the field interaction) as a generic quantity *g*. (The basic points carry over to the detailed treatment with spinors.) In accordance with a common convention, time advances from bottom to top in the diagrams; electron lines are denoted with arrows in the advancing time direction and positron with reversed arrows; photon lines are wavy. Key components of the Feynman amplitudes for each process are:

(i) incoming, external, 'free' particle lines of momentum $p_j$, labeled by *exp[-ip_j.x_i]*, i,j=1,2[13]

(ii) outgoing, external, 'free' particle lines of momentum $p_k$, labeled by *exp[ip_k.x_i]*, k=3,4

(iii) coupling amplitudes *ig* at each vertex

(iv) an internal 'virtual' photon line of (variable) momentum *q*, labeled by the generic propagator[14]

---

[13] These plane waves are simplified components of the currents appearing in (1) and (2).

[14] The term 'generic' reflects the fact that the denominator here is simply $q^2$. The different types of propagators involve different prescriptions for the addition of an infinitesimal imaginary quantity, for dealing with the poles corresponding to 'real' photons with $q^2=0$. However, in actual calculations, one

$$D(x-y) = \int \frac{d^4q}{(2\pi)^4} \frac{ie^{iq\cdot(x_1-x_2)}}{q^2} \quad \text{(for } m=0 \text{ in the photon case)}. \tag{3}$$

To calculate the amplitude applying to the first diagram, these factors are multiplied together and integrated over all spacetime coordinates $x_1$ and $x_2$ to give an amplitude $M_1$ for the first diagram. Specifically:

$$M_1 \propto \int d^4x_1 d^4x_2 e^{-ip_1\cdot x_1} e^{-ip_2\cdot x_2} (ig) \frac{d^4q}{(2\pi)^4} \frac{ie^{iq\cdot(x_1-x_2)}}{q^2} (ig) e^{ip_3\cdot x_1} e^{ip_4\cdot x_2} \tag{4}$$

The integrations over the spacetime coordinates $x_i$ yield delta functions imposing conservation of energy at each vertex (which are conventionally disregarded in subsequent calculations). A similar amplitude analysis applies to the second diagram, giving $M_2$. Then the two amplitudes for the two diagrams are summed, giving the total amplitude $M$ for this scattering process. $M$ (a complex quantity) is squared to give the probability of this particular scattering process: $P(p_1,p_2 \to p_3,p_4) = M^*M$. This is the probability of observing outgoing electron momentum $p_3$ and positron momentum $p_4$ given incoming electron and positron momenta of $p_1$ and $p_2$ respectively. It is interesting to note that the amplitude for the (lowest order) scattering process is the *sum* of the two diagrams in Figure 2, meaning that each is just an offer wave component and that the two mutually interfere (see also Figure 3).

3.2 "Free" particles vs. "virtual" particles

Now, for our purposes, the thing to notice is that, in this very typical analysis, we disregard the history of the incoming particles and the fate of the outgoing particles. They are treated in the computation as 'free' particles—particles with infinite lifetimes— whether this is the case or not. And it actually can't be, since we have prepared the

---

often simply uses this expression. The fact that the generic expression yields accurate predictions can be taken as an indication that the theoretical considerations surrounding the choice of propagator do not have empirical content in the context of micro-processes such as scattering.

incoming particles to have a certain known energy and we detect the outgoing particles to see whether our predictions are accurate. We simply exclude those emission and detection processes from the computation because it's not what we are interested in. We are interested in a prediction *conditioned* on a certain initial state and a certain final state. This illustrates how the process of describing and predicting an isolated aspect of physical reality necessarily introduces an element of distortion in that it misrepresents those aspects not included in the analysis (i,e,., misrepresents 'virtual' photons—i.e., photons with finite lifetimes-- as 'real' photons). This is perhaps yet another aspect of the riddle of quantum reality in which one cannot accurately separate what is being observed from the act of observation: the act of observation necessarily distorts, either physically or epistemologically (or both), what is being observed.

3.3. The PTI account of scattering

Now, let us see how PTI describes the scattering process described above. There is a two-particle offer wave, an interaction, and a detection/absorption. The actual interaction encompasses all orders[15]---not just the lowest order interactions depicted here---so the initial offer wave becomes fractally articulated in a way not present in the nonrelativistic case. The fractal nature of this process is reflected in the perturbative origin of the S-matrix, which allows for a theoretically unlimited number of finer and more numerous interactions.[16] All possible interactions of a given order, over all possible orders, are superimposed in the relativistic offer wave corresponding to the actual amplitude of the process. (Herein we gain a glimpse of the astounding creative complexity of Nature. In practice, only the lowest orders are actually calculated; higher order calculations are simply too unwieldy, but excellent accuracy is obtained even restricted to these low orders.)[17]

---

[15] To be precise, all orders up to a natural limit short of the continuum; see footnote 18.

[16] I refer here to the distinguishing features of fractals: (1) a progressively finer structure continuing to arbitrarily small scales and (2) self-similarity in the 'branching' of those finer processes from the 'parent' process.

In the standard approach, this final amplitude is squared to obtain the probability of the corresponding event, but the squaring process has no physical basis—it is simply a mathematical device (the Born Rule). In contrast, according to PTI, the absorption of the offer wave generates a confirmation (the 'response of the absorber'), an advanced field. This field can be consistently reinterpreted as a retarded field from the vantage point of an 'observer' composed of positive energy and experiencing events in a forward temporal direction. The product of the offer (represented by the amplitude) and the confirmation (represented by the amplitude's complex conjugate) corresponds to the Born Rule.[18] This quantity describes, as in the non-relativistic case, an incipient transaction reflecting the physical weight of the process. In general, other, 'rival' processes will generate rival confirmations (for example, the detection of outgoing particles of differing momentum) from different detectors and will have their own incipient transactions. A spontaneous symmetry-breaking occurs,[19] in which the physical weight functions as the probability of that particular process as it 'competes' with other possible processes. The final result of this process is the actualization of a particular scattering event (i.e., a particular set of outgoing momenta) in spacetime.

---

[17] Adopting a realist view of the perturbative process might be seen as subject to criticism based on theoretical divergences of QFT; i.e., it is often claimed that the virtual particle processes corresponding to terms in the perturbative expansion are 'fictitious.' But such divergences arise from taking the mathematical limit of zero distances for virtual particle propagation. This limit, which surpasses the Planck length, is likely an unwarranted mathematical idealization. In any case, it should be recalled that spacetime indices really characterise points on the quantum field rather than points in spacetime (Auyang 1995, 48); according to PTI, spacetime emerges only at the level of actualized transactions. Apart from these ontological considerations, progress has been made in discretised field approaches to renormalization such as that pioneered by Kenneth Wilson (lattice gauge theory, cf. Wilson 1971, 1974, 1975). Another argument against the above criticism of a realist view of QFT's perturbative expansion is that formally similar divergences appear in solid state theory, for example in the Kondo effect (Kondo, 1964), but these are not taken as evidence that the underlying physical model should be considered 'fictitious.'

[18] Technically, by comparison with the standard time-asymmetric theory, the product of the original offer wave component amplitude, ½ $a$, and its complex conjugate, ½ $a^*$ |, yields an overall factor of ¼, but this amounts to a universal factor which has no empirical content since it would apply to all processes and therefore would be unobservable.

Thus, upon actualization of a particular incipient transaction, this confirmation *adds* to the offer and provides for the unambiguous propagation of a full-strength, positive-energy field in the t >0 direction and cancellation of advanced components; this is essentially the process discussed by Davies, above, in which the Earth-Moon energy propagation must be described by $D_F$ rather than by $\overline{D}$.

3.4 Internal couplings and confirmation in relativistic PTI

Now we come to the important point introduced in the Abstract and in §1.1. Notice that the internal, unobserved processes involving the creation and absorption of virtual particles, are not considered as generating confirmations in relativistic PTI (see Figure 3.) These are true 'internal lines' in which the direction of propagation is undefined; therefore $D_F$ can be replaced by $\overline{D}$. These must not be confirmed, because if they were, each such confirmation would set up an incipient transaction and the calculation would be a different one (i.e., one would not have a sum of partial amplitudes $M_1$ and $M_2$ *before* squaring; squaring corresponds to the confirmation). This situation in which several (in principle, an infinite number of)[20] offer wave components are summed to obtain the total scattering offer wave) involves field coupling amplitudes, which are not present in the non-relativistic case. The internal coupling amplitude represents a kind of counterfactual situation; namely, *the amplitude for a confirmation to be generated in the case when one was not in fact generated*. This is a novel feature of the interpretation

---

[19] The symmetry breaking aspect of transaction actualization is introduced in Kastner (2011 c) and further explored in a forthcoming work.

[20] For the present argument, I disregard the issue of renormalization, in which an arbitrary cutoff is implemented in order to avoid self-energy divergences resulting from this apparently infinite regression. But see note 18 for why the latter is probably a mathematical idealization not applicable to physical reality.

appearing only at the relativistic level, in which the number and type of particles can change.

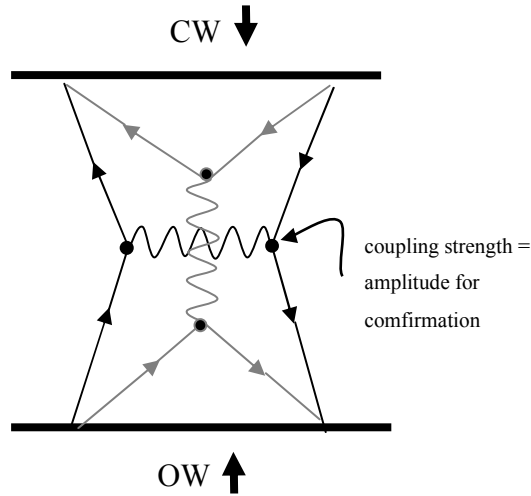

Figure 3. Both diagrams of Figure 2 are actually superimposed in calculating the amplitude for the offer wave corresponding to Bhabha scattering ($M_1$ shown in black and $M_2$ shown in grey). Confirmations occur only at the external, outgoing ends. Coupling amplitudes at vertices are amplitudes for confirmations that did not, in fact, occur in this process but must still be taken into account in determining the probability for the event. There is no 'fact of the matter' about which of the scattering processes was followed, just as there is no fact of the matter about which slit a particle took in the two-slit experiment with interference.

The physical meaning of the coupling amplitude involves a subtle conceptual step, so let us examine this in further detail. An amplitude less than unity for a confirmation, as applied to an internal vertex in calculating a scattering amplitude, takes into account that the confirmation did not occur for the present situation but that there was a possibility for it to have occurred. Thus the interaction of the incoming electron state with the virtual photon can be thought of as the interaction of an electron OW with a

*potential* absorber.[21] The condition for a relativistic scattering process is that a confirmation was *possible,* even though it did not in fact occur for that process. We know the confirmation did not occur because the photon is only a virtual one, but we have to allow for its possibility because that is what constitutes the interaction at any vertex. Considering the possibility of a confirmation to be strictly zero corresponds to the vanishing of the vertex and therefore the vanishing of the scattering process under consideration. The element of stochasticity accompanying the coupling amplitude is distinct from the stochastic nature of the realization of one from a set of competing incipient transactions. The stochasticity of the internal coupling amplitude constitutes yet a more subtle, purely relativistic level of possibility beneath the nonrelativistic situation in which CW have *de facto* been generated.

Concerning the relationship of the amplitude for confirmation to the probability of confirmation: recall that the probability of the scattering process under consideration is defined by the product of the OW amplitude for the whole process times the CW amplitude for the whole process. The latter arises from a confirmation generated for the outgoing two-particle state. A necessary (but not sufficient) condition for the given scattering process to occur is that the CW for the outgoing state be generated, and that CW will pick up additional factors of all coupling amplitudes[22]. Thus the coupling amplitudes will be accompanied by adjoint factors, giving the relevant unit of probability. But again, *coupling amplitudes* are fundamentally different from the amplitudes of a given OW or CW. The former describe an uncertainty in the generation of the OW or CW itself[23] while the latter describe the amplitude of an existing OW or CW. The former does not arise at the nonrelativistic level but appears in the relativistic domain as a still more tenuous form of physical possibility than the OW and CW.

---

[21] The photon, being virtual, cannot attain full absorber status; the other way to see this is that conservation of energy does not allow a single photon to absorb a single electron.

[22] Since those expressions will in general be sums, there will be cross terms mixing the couplings, but this simply illustrates the higher level of interference in the relativistic case.

[23] The discussion has addressed coupling amplitudes as amplitudes for confirmations, but a vertex also involves the possible generation of an OW. One can think of the virtual particle as the manifestation of *both* the 'failed' OW and CW.

To understand this novel feature of the relativistic PTI, consider (as a rough analogy) a coin toss. The nonrelativistic situation is comparable to a coin toss in which the only possibilities are heads (no CW is possible) or tails (a CW is definitely generated). Now imagine that you have a very thick coin in which it is possible to have the coin land on its edge: this represents the relativistic situation. The result is neither heads nor tails but something else entirely, which can lead to a further situation not available in the two-valued case: namely, the exchange of one or more virtual particles and outgoing states differing from the incoming states.

Does this make the idea of 'absorber' arbitrary? No; it simply means that one can only know after the fact that a CW has been generated. When a CW is in fact generated, the generating entity is unambiguously an absorber. In the nonrelativistic case we deal only with situations in which CW did in fact occur. Section 5 discusses in more detail the implication of this picture for the microscopic/macroscopic boundary.

A rough analog of the scattering situation can be found in the case of a quantum in the two-slit experiment. For a particle created at source S, passing a screen with two slits A and B, and being detected at position X on a final screen, the partial amplitudes are

$$<X|A><A|S> \qquad (5a)$$

$$<X|B><B|S>. \qquad (5b)$$

These must be added together to obtain the correct probability for detection at point X, yet neither generates a confirmation (if both slits are open and there are no detectors at the slits). The confirmations in play are those arising from the possible screen positions $X_i$ on the final screen, and those propagate back through both slits. In each case, no particle was detected at either slit, but the existence of the slit[24] requires that we take it into account. In the same way, the couplings at vertices resulting in virtual, intermediate quanta represented in the Feynman diagrams must be taken into account. These vertices are analogous to the slits in the two-slit experiment, and the confirmations propagating back through both possible scattering processes are those generated at the outgoing ends of the scattering process for different possible outgoing particle states.

In quantum mechanics, the unobservable must be accounted for, and it is accounted for in terms of amplitudes (partial offers and partial confirmations), not in terms of probabilities. In the two-slit experiment, the partial confirmations are the advanced wave components from point X on the final screen, through the slits, to the source: <S|A><A|X> and <S|B><B|X>. In the scattering example, the partial confirmations are the advanced wave components for given outgoing particle states, through each possible scattering process, to the initial particle states. The disanalogy between the two cases consists in the fact that the two individual 'which slit' components exist in the absence of any possible intermediate confirmation, while in the scattering example the two individual scattering amplitudes pictured in Figure 2 exist only in virtue of the possible internal, counterfactual confirmations defining the vertices.

4, Classical limit of the quantum electromagnetic field

It is interesting and instructive to consider how the classical Wheeler-Feynman theory can be seen as a limit of the quantized version. In this section I show how the classical, real electromagnetic field emerges from the domain of complex, pre-empirical offer and confirmation waves that are ontologically distinct from classical fields.

---

[24] To be more precise in terms of TI, the existence of a large number of absorbers (the slitted screen) which allow only specific OW components to proceed through the experiment.

It first needs to be kept in mind that a classical field $E(x,t)$ assumed to propagate in spacetime is replaced by an operator $\hat{E}(x,t)$ in the context of relativistic quantum theory; the latter is a very different entity. It is the transition amplitude of a product of such field operators (actually the vector potential, $\hat{A}(x,t)$ [25]) corresponding to two different states of the field (or spacetime points)[26] which then replaces the classical propagating field. That quantity (also known as the Feynman propagator $D_F$, discussed above, when constructed to ensure that only positive energies are directed toward the future) is now a probability amplitude only, and thus corresponds to the offer wave (OW) component of nonrelativistic PTI.

Let us now consider how the classical electromagnetic field emerges from the quantum theoretic electromagnetic field by way of the transactional process. In order to do this, it must first be noted that so-called 'coherent states' $|\alpha\rangle$ of quantum fields provide the closest correspondence between these and their classical counterparts. Such states have an indeterminate number of quanta such that annihilation (detection/absorption) of any finite number of quanta does not change the state of the field:

$$|\alpha\rangle = e^{\frac{-|\alpha|^2}{2}} \sum_{n=0,\infty} \frac{\alpha^n}{\sqrt{n!}} |n\rangle \qquad (10)$$

These states are eigenstates of the field annihilation operator $\hat{a}$; the field in that state does not 'know' that it has last a photon. That is,

---

[25] The electromagnetic field and the electromagnetic vector potential are related by

$$\vec{E}(x,t) = -\frac{1}{c}\frac{\partial \vec{A}}{\partial t} - \nabla A_t .$$

[26] In practice, when the initial and final states are spacetime points, they are 'dummy variables'; i.e., variables of integration. In quantum field theory it is not meaningful to talk about a quantum being created and destroyed at specific spacetime points.

$$\hat{a}|\alpha\rangle = \alpha|\alpha\rangle; \qquad (11)$$

so that it has an effectively infinite and constantly replenished supply of photons. The coherent state can be thought of as a 'transaction reservoir' analogous to the temperature reservoirs of macroscopic thermodynamics. In the latter theory, the interaction of a system of interest with its environment is modeled as the coupling of the system to a 'heat reservoir' of temperature T. In this model, exchanges of heat between the reservoir and the system affect the system but have no measurable effect on the reservoir. In the same way, a coherent state is not affected by the detection of finite numbers of photons.

Experiments have been conducted in which a generalized electromagnetic field operator is measured for such a state.[27] Detections of photons in the coherent field state generate a current, and that current is plotted as a function of the phase of the monochromatic source (i.e., a source oscillating at a particular frequency—for example, a laser). (See Figure 6) Such a plot reflects the oscillation of the source in that the photons are detected in states of the measured observable (essentially the electric field amplitude) which oscillate as a function of phase (individual photons do not oscillate, however).

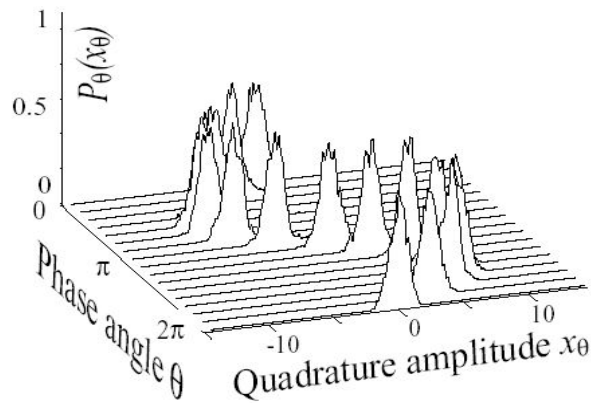

---

[27] See, for example, Breitenbach, Schiller and Mlynek (1997).

Figure 6 : Data from photon detections reflecting oscillation of the field source.[28]

The theoretical difference between the quantum versions of fields (such as the coherent state) and their classical counterparts can be understood in terms of the ontological difference between quantum possibilities (offer and confirmation waves and incipient transactions) and structured sets of actualized transactions. The quantized fields represent the creation or destruction of possibilities, and the classical fields arise from states of the field that sustain very frequent actualized transactions, in which energy is transferred essentially continuously from one object to another. Again, this can be illustrated by the results of experiments with coherent states that 'map' the changing electric field in terms of photon detections, each of which is a transaction. For states with small average photon numbers, the field amplitude is small and quantum 'noise' is evident (for the coherent state, these are the same random fluctuations found in the vacuum state). As the coherent state comprises larger and larger numbers of photons, the 'signal to noise ratio' is enhanced and it approaches a classical field (see Figure 7.) Thus the classical field is the quantum coherent state in the limit of very frequent detections/transactions.

---

[28] Figures 4 and 5 are reproduced from the dissertation of Breitenbach and are in the public domain at http://upload.wikimedia.org/wikipedia/commons/1/1a/Coherent_state_wavepacket.jpg

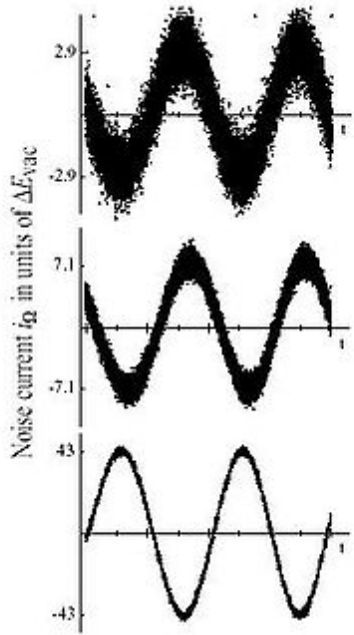

Figure 7: Coherent states with increasing average photon number (top to bottom).

It is the classical, continuous detection/transaction limit, in which the field can be thought of as a classical propagating wave, to which the original Wheeler-Feynman theory applies. But it is important to keep in mind the fundamental distinction between a classical field and its quantum counterpart. In this regard, Paul Dirac has observed that

"Firstly, the light wave is always real, whereas the de Broglie wave associated with a light quantum moving in a definite direction must be taken to involve an imaginary exponential. A more important difference is that their intensities are to be interpreted in different ways. The number of light quanta per unit volume associated with a monochromatic light wave equals the energy per unit volume of the wave divided by the energy $h\nu$ of a single light quantum. On the other hand, a monochromatic de Broglie wave of amplitude $a$ (multiplied into the imaginary exponential factor) must be interpreted as representing $a^2$ light-quanta per unit volume for all frequencies." (Dirac 1927, p. 247)

Dirac's comments highlight the ontological distinction between the classical electromagnetic wave and the quantum state (de Broglie wave) for the electromagnetic field. Whereas the classical wave conveys energy through its intensity (the square of its electric field strength), the quantum wave conveys possibility—that is, its square conveys probability in that it represents an incipient transaction whose weight corresponds (in nonrelativistic quantum mechanics) to the probability of the corresponding event; or, in the case of a coherent field state, the *number of quanta* most likely to be actualized. The amplitude of a de Broglie wave for a coherent state with average photon number N is equal to $\sqrt{N}$ (which is proportional to the electric field amplitude for the state); it is a multi-quantum probability amplitude that , when squared, predicts that the most probable number of photons to be detected will be N. Thus, if a coherent state with average photon number 3 were enclosed in a perfectly absorbing box, on examining the box after a time period significantly greater than the inverse frequency of the field (i.e., the period of the oscillation), it would ideally be found to have detected 3 photons.

One could do this by measuring the energy increase of the box, but that is not required; one could imagine a box constructed out of photographic plates that could provide images (dots) of photon absorption. Such images provide a simple numerical answer to the question: "How many photons were actualized?"—and it is this question to the squared amplitude ($|\alpha|^2$) of the coherent state $|\alpha\rangle$ applies. In contrast, the squared amplitude of the classical wave addresses the question, "what is the energy associated with the actualized photons?" The *energy $E = h\nu$* of a particular actualized (detected) photon is frequency-dependent, but the *probable number* of actualized photons is not.

Yet the unity of the two descriptions is still expressed in the fact that it is not the classical field that  really conveys energy: rather, it is the *intensity* (squared amplitude) of the field. This can again be traced to the underlying transactional description. A photon does not exist in spacetime unless there is an actualized transaction involving an offer wave and a confirmation wave, which is what is described by the squaring process (Born

Rule).[29] Energy can only be conveyed by a detected photon, not by an amplitude (offer wave) only. This fact appears at the classical level and can be seen as a kind of 'correspondence principle' between the two descriptions.

5. Concluding remarks

The new Possibilist Transactional Interpretation has been extended to the relativistic domain. It has been argued that PTI is fully compatible with relativity. Moreover, extending PTI in this way breaks new ground in the understanding of 'absorber' in the transactional picture. It has been proposed herein that the coupling amplitudes between interacting fields in the relativistic domain are to be identified as the amplitudes for the generation of confirmation waves. This means that the generation of CW is a fundamentally stochastic process. Thus, although PTI can provide a definitive physical account of measurement, that does not translate into a causal, predictable account of measurement. At the relativistic level, there is no way to predict with certainty whether, for any given instance, a CW will be generated. Couplings between (relativistic) quantum fields give only amplitudes for confirmations.

Thus, we can only know after the fact that a CW has been generated. However, in macroscopic situations, detectors are composed of huge numbers of individual potential absorbers, and this virtually assures the generation of CW somewhere in the detector. In fact, the identification of coupling amplitudes as amplitudes for confirmations allows us to specify an unambiguous physical basis for the notorious 'micro/macro' boundary. The macroscopic world is simply the level at which CW are virtually assured. This prevents the propagation of quantum superpositions to the observable level (i.e., it prevents a macroscopic object from being in a superposition), because it is overwhelmingly likely that a CW will be generated before the point that, say, a cat would need to be described by a superposition of states of its physical health.

---

[29] For those concerned about whether the universe may not be a 'light tight box' as required by traditional 'direct action' (DA) theories, thus not providing for full future absorption of the OW, it should be noted that confirmations may also be provided by a perfectly reflecting past boundary condition, as proposed in Cramer (1983). This is a type of 'absorberless' confirmation in which the advanced wave from the emitter is reflected at $t=0$ and thereby cancels the remnant advanced wave from the emitter and builds the emitter's retarded OW up to full strength, resulting in an actualized transfer of energy into the infinite future.

For example, suppose we want to detect photons by way of the photoelectric effect (in which electrons are ejected from the surface of a metal by absorbing the energy of an incoming photon). In this case, the electron serves as the potential absorber for the photon's OW. The coupling of any individual photon and electron has an amplitude of ~ 1/137, about .007, so the probability that an interaction between the two will generate a CW is ~ $(.007)^2$ = .00005, a very small number. Indeed the probability that a CW will *not* be generated in an interaction of a photon with a single electron is 1- .00005 = .99995, so for any individual electron we may confidently predict that a CW will *not* be generated. However, consider a small macroscopic sample of metal, say 1 cubic centimeter. A typical metal has roughly $10^{23}$ free (conduction) electrons in this size sample. The probability that a CW will *not* be generated in the sample is the product of the probabilities of *every* electron in the sample not generating a CW, or (.99995) raised to the $10^{23}$ power, which is an infinitesimal number (your calculator will give zero). Thus it is virtually certain that in any interaction of a photon with a macroscopic quantity of electrons, at least one of them will generate a CW. So when we deal with a macroscopic detector, it is virtually certain that an OW will generate a 'response of the absorber' somewhere in the detector, thus physically warranting the Born Rule's squaring procedure.

The preceding explains why we have to work so hard to retain quantum superpositions of objects. In order to obtain observable phenomena we must work with macroscopic quantities of matter, but this is the level at which confirmations (absorber responses) become virtually certain, and the latter are what cause the 'collapse' of superpositions of quantum states[30]. With increasing technological sophistication comes the ability to create superpositions of mesoscopic objects such as "Buckeyballs" (cf. Arndt and Zeilinger, 2003). As noted above, it is certainly possible for a confirmation wave to be generated by a single electron. We usually don't notice such situations because they would not be selected as 'applicable runs' in an experiment looking, for example, at photon diffraction.

---

[30] By 'collapse of superpositons', I refer to the superposition of a single object. For example, we can see evidence of superposition in the two-slit experiment, in which both the OW and CW have access to both slits, but any individual mark corresponding to a particular 'hit' is itself in a determinate position (i.e., not in a superposition of positions).

Finally, an interpretational dividend of this relativistic version of PTI is that it provides for a solution to the measurement problem by pointing out that the latter cannot be solved at the nonrelativistic level. That is, in order to define 'measurement' in physical terms, one must take into account the physical process of absorption, which is fundamentally a relativistic process. This fact not only underscores the power of PTI as an interpretation; it also explains why the measurement problem has not been convincingly solved in extant competing interpretations of nonrelativistic quantum mechanics less amenable to relativistic extension.

Acknowledgements

I am grateful to two anonymous reviewers for their helpful comments in improving the presentation of this paper.